\documentclass[9pt,conference]{IEEEtran}
\usepackage[T1]{fontenc}
\usepackage[latin9]{inputenc}
\usepackage{geometry}
\usepackage{color}
\usepackage{units}
\usepackage{amsmath}
\usepackage{amssymb}
\usepackage{graphicx}

\makeatletter
\IEEEoverridecommandlockouts

\@ifundefined{showcaptionsetup}{}{%
 \PassOptionsToPackage{caption=false}{subfig}}
\usepackage{subfig}
\makeatother

\begin{document}

\title{Multi-Path TCP with Network Coding for Mobile Devices in Heterogeneous
Networks\thanks{This work is sponsored, in part, by the Assistant Secretary of Defense (ASD R\&E) under Air Force Contract \# FA8721-05-C-0002.  Opinions, interpretations, recommendations and conclusions are those of the authors and are not necessarily endorsed by the United States Government.}\textcolor{black}{\vspace{-5pt}}}

\author{{\normalsize Jason Cloud$^{\dagger\ast}$, Flávio du Pin Calmon$^{\dagger\ast}$,
Weifei Zeng$^{\dagger}$, Giovanni Pau$^{\ddagger}$, Linda M. Zeger$^{**}$,
and Muriel Médard$^{\dagger}$}\\
{\normalsize $^{\dagger}$Research Laboratory of Electronics, Massachusetts
Institute of Technology, Cambridge, MA}\\
{\normalsize $^{\ddagger}$Computer Science Department, University
of California, Los Angeles, CA}\\
{\normalsize $^{\ast}$MIT Lincoln Laboratory, Lexington, MA}\\
{\normalsize $^{**}$Auroral LLC}\\
{\normalsize Email: \{jcloud, flavio, weifei\}@mit.edu, gpau@cs.ucla.edu,
zeger@auroral.biz, medard@mit.edu}\textcolor{black}{\vspace{-15pt}}}
\maketitle
\begin{abstract}
Existing mobile devices have the capability to use multiple network
technologies simultaneously to help increase performance; but they
rarely, if at all, effectively use these technologies in parallel.
We first present empirical data to help understand the mobile environment
when three heterogeneous networks are available to the mobile device
(i.e., a WiFi network, WiMax network, and an Iridium satellite network).
We then propose a reliable, multi-path protocol called Multi-Path
TCP with Network Coding (MPTCP/NC) that utilizes each of these networks
in parallel. An analytical model is developed and a mean-field approximation
is derived that gives an estimate of the protocol's achievable throughput.
Finally, a comparison between MPTCP and MPTCP/NC is presented using
both the empirical data and mean-field approximation. Our results
show that network coding can provide users in mobile environments
a higher quality of service by enabling the use of multiple network
technologies and the capability to overcome packet losses due to lossy,
wireless network connections.
\end{abstract}

\section{Introduction\label{sec:Introduction}}

Simultaneous use of multiple network interfaces on a single mobile
device has the potential to increase quality of service, seamlessly
offload traffic from expensive networks to cheaper ones, increase
session reliability, etc.; yet current technology does not utilize
the available resources efficiently to meet these objectives. Instead,
only a single network interface is preferred while the others are
left unused. For example, consider a standard smart phone that has
a cellular data connection, such as 3G or LTE, and a WiFi connection.
Data is sent over either one or the other, but not both. Given that
existing infrastructure currently supports the use of both WiFi and
cellular technologies (\cite{chen_network_2012}, \cite{sommers_cell_2012}),
new techniques must be developed to properly leverage all available
resources, regardless of their quality, to increase mobile user performance.

A significant amount of research has been performed that attempts
to utilize these heterogeneous network connections. For example, Multi-Path
TCP (MPTCP) is a new protocol currently in the working group level
of the IETF \cite{mptcpietf}. The protocol adds a new layer above
the transport layer which provides packet scheduling across multiple
TCP sub-flows and guarantees packet delivery through the use of a
somewhat complex management scheme. Furthermore, MPTCP uses TCP as
its primary flow control mechanism on each of the sub-flows. While
the use of TCP ensures fairness with other TCP flows, the performance
of TCP over lossy networks (e.g., wireless networks) is known to be
poor \cite{xylomenos2001tcpperformance}. Network coding is one possible
solution that both reduces the need for a complex management scheme
and can increase TCP's performance over lossy networks.

Several suggestions on how to incorporate network coding with MPTCP
have been proposed. Gheorgiu \textit{et. al.} \cite{gheorghiu2010multipath}
propose a protocol called CoMP that uses network coding for multi-path
transmission that incorporates only some aspects of TCP. \cite{xiazhuoqun2009animproved}
and \cite{zhuo-qunxia2009amultipath} add a multi-path scheduler below
the TCP, network coding, and IP layers negating the congestion control
benefits of TCP over single paths. Finally, ParandehGheibi \textit{et.
al.} \cite{parandehgheibi2010accessnetwork}, and implemented by \cite{kulkarni2011animplementation}
in OpNet, provide a sub-flow selection control policy for network
coded packets over heterogeneous networks that optimizes the trade-offs
between the network usage costs and the Quality of user Experience
(QoE) for media-streaming applications. Many approaches have also
been proposed to increase TCP's performance in the presence of high
losses (\cite{liu2001atcptcp}, \cite{caceres1995improving}, \cite{kaixu2004tcpjersey},
\cite{chandran2001afeedbackbased} to name a few). One promising approach
is TCP/NC proposed by Sundararajan \textit{et. al.}, \cite{sundararajan2011network}.
TCP/NC introduces a layer between TCP and IP that uses random linear
network coding \cite{ho2006arandom} to produce linear combinations
of all packets contained in the TCP congestion control window. These
coded packets are then transmitted over the network and decoded by
a client. As shown in \cite{sundararajan2011network}, network coding
helps to alleviate the effects of packet loss due to poor channels
while preserving the congestion control and fairness mechanisms provided
by TCP. 

In this paper, we first present empirical measurements for the simultaneous
use of three heterogeneous network connections (e.g., WiFi, WiMax,
and an Iridium satellite network) in a mobile environment. These measurements
highlight the fact that none of the networks provide 100\% reliable
communication; but in combination, the simultaneous use of these networks
can provide significant gains over the use of only one at a time.
We then present a model based on \cite{padhye2000modeling} and \cite{kim2011modeling},
as well as derive a mean-field approximation for the throughput of
both MPTCP and MPTCP with network coding (MPTCP/NC). MPTCP/NC uses
network coding prior to packet sub-flow scheduling to simplify the
MPTCP management scheme, and uses network coding a second time below
TCP to provide a mechanism to overcome packet losses induced by lossy,
wireless networks. We conclude with a comparison of MPTCP and MPTCP/NC
using both the mean-field approximation and the experimentally collected
data to show that network coding can provide beneficial enhancements
to the existing protocols.

The remainder of the paper is organized as follows. In Section \ref{sec:Empirical-measurements},
we outline the experimental measurement setup and present an overview
of the collected data. Section \ref{sec:Analytical-Models} provides
a brief description of MPTCP and MPTCP/NC, develops the analytical
models used, and derives the mean-field approximations for both protocols.
We then compare the performance of both MPTCP and MPTCP/NC in Section
\ref{sec:Performance} and conclude in Section \ref{sec:Conclusion}.

\section{Empirical Measurements\label{sec:Empirical-measurements}}

Using a WiMax base station, a WiFi mesh network, and an Iridium satellite
data modem \cite{iridium}, simultaneous network traces were collected
between the Network Research Laboratory (NRL), Department of Computer
Science, UCLA and a vehicle driving a fixed route around the UCLA
campus. Each experiment sent packets, varying between 64 bytes, 512
bytes , and 1,350 bytes in size, at rates based on the direction of
travel. For example, traffic generated by the computer in the NRL
and sent to the vehicle, referred to as downlink (D/L) traffic, was
sent at rates determined by the individual network (WiMax: 20 Mbps,
WiFi: 20 Mbps, and Iridium: 1 kbps). Traffic generated by the computer
in the vehicle and sent to the computer in the NRL, referred to as
uplink (U/L) traffic, was also sent at rates determined by the individual
network (WiMax: 1 Mbps, WiFi: 20 Mbps, and Iridium: 1 kbps). In each
experiment, only D/L traffic or U/L traffic was generated.\textcolor{black}{\vspace{-3pt}}

\subsection{Testbed Configuration\label{sub:Testbed-Configuration}}

\begin{figure}
\begin{centering}
\subfloat[Network Configuration]{\begin{centering}
\includegraphics[width=0.45\columnwidth]{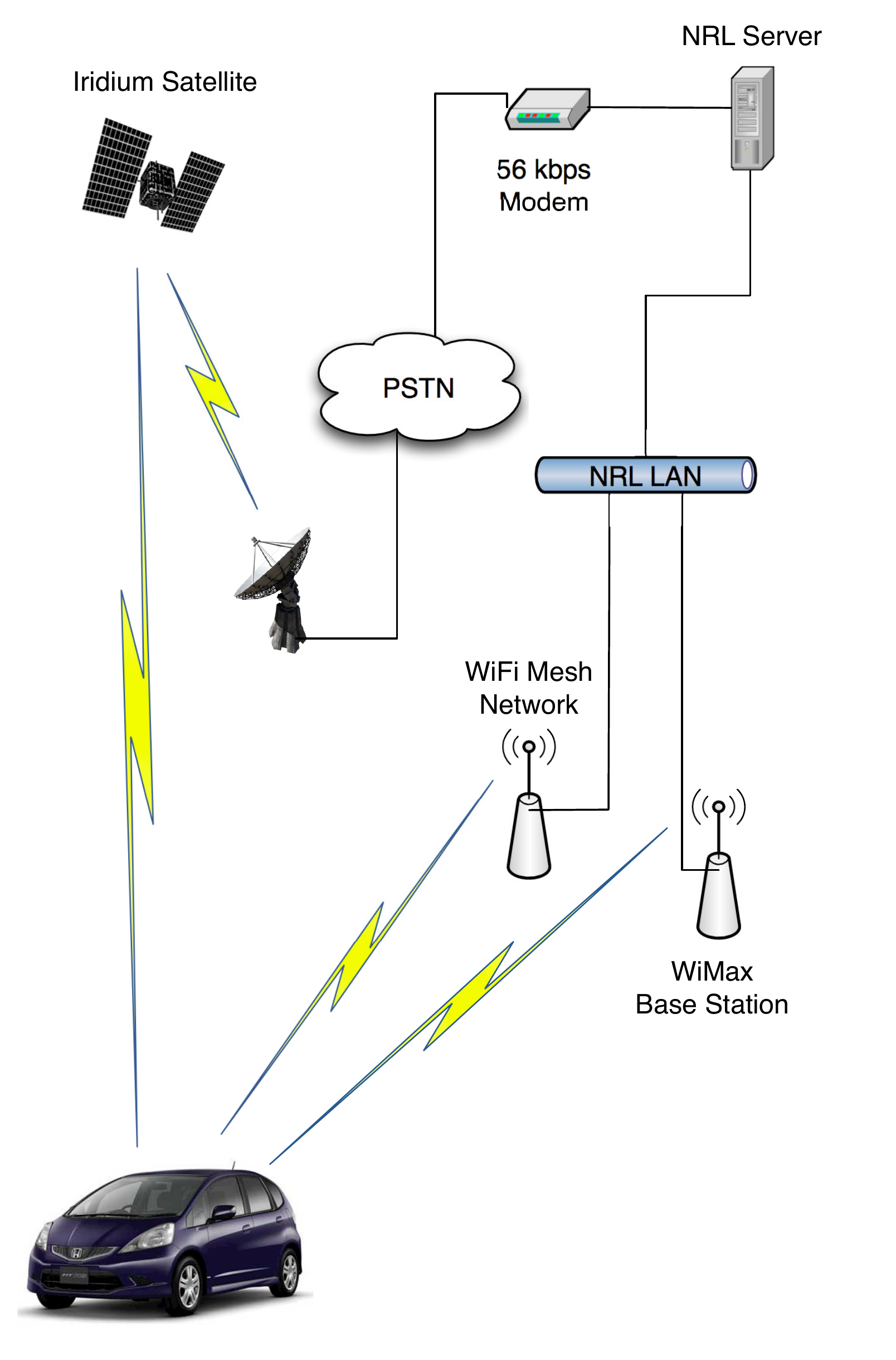}
\par\end{centering}

}\subfloat[WiMax/WiFi base station placement and vehicle route.]{\begin{centering}
\includegraphics[width=0.5\columnwidth]{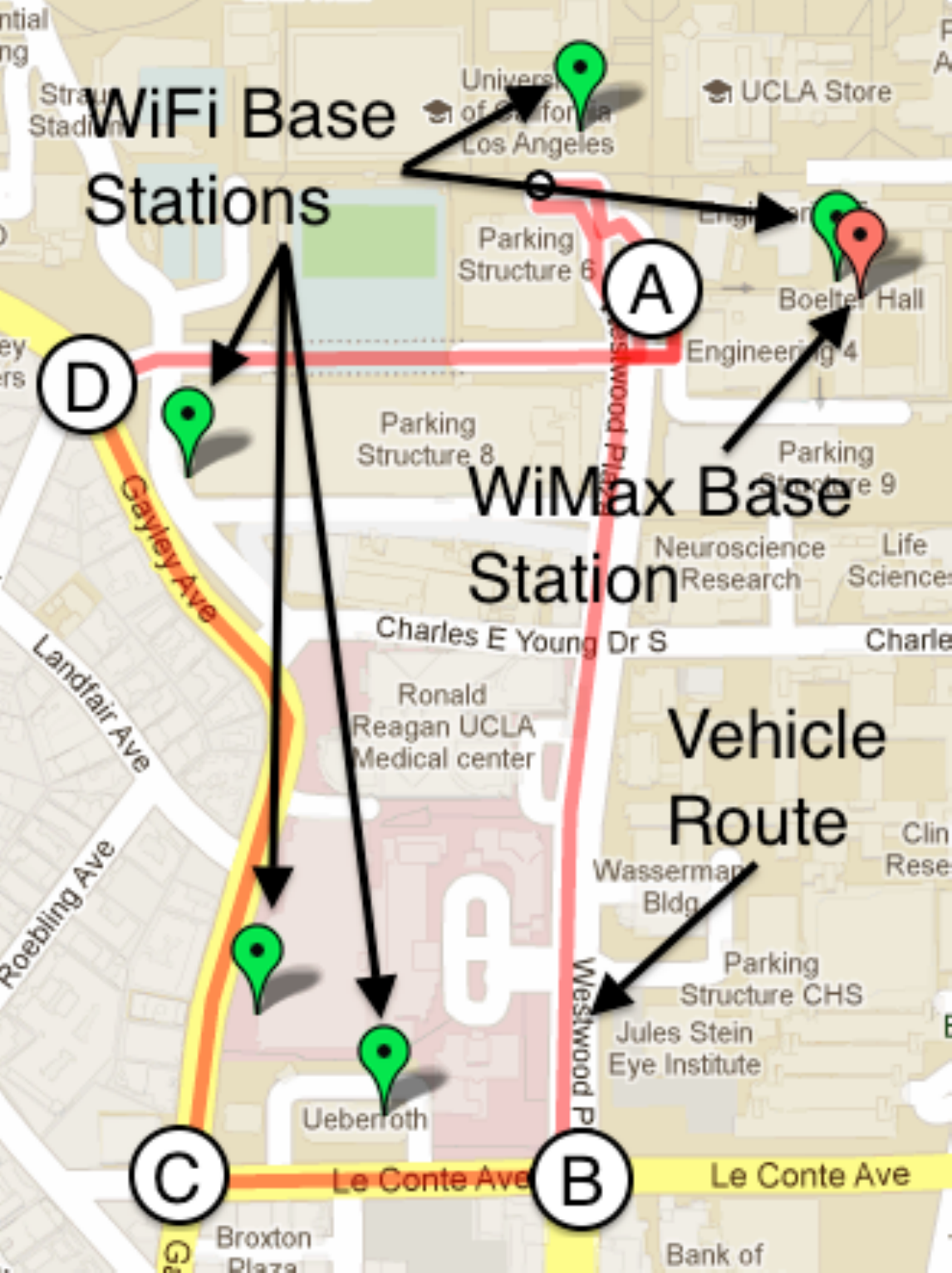}
\par\end{centering}

}
\par\end{centering}

\caption{Empirical measurement collection setup.\label{fig:Test-Setup}}
\textcolor{black}{\vspace{-20pt}}
\end{figure}

Measurements were taken between a mobile commodity laptop and a fixed
server located within the NRL. The computer in the NRL was connected
to the NRL LAN which has gateways to both the WiMax base-station and
WiFi mesh network. A 56 kbps modem was used to connect the computer
to the public switched telephone network (PSTN) in order to utilize
the Iridium satellite network. In the vehicle, a single computer with
separate WiMax and WiFi cards, as well as a connection to an Iridium
data modem, was used to transmit and receive data. A diagram of the
setup is shown in Figure \ref{fig:Test-Setup}(a). UDP network traffic
was generated using Iperf \cite{iperf} and network traces were collected
using tshark (a command line version of Wireshark) \cite{tshark}
on both the computers. 

The vehicle containing the mobile computer travelled a fixed route
through the UCLA campus chosen so that the vehicle passed in and out
of the coverage areas of all three networks. For example, connections
through all three networks was established prior to each experiment.
The vehicle would then drop from and reconnect to each of the individual
networks, depending on the location of the vehicle and coverage of
the specific network, throughout the duration of the experiment. Figure
\ref{fig:Test-Setup}(b) provides the vehicle route and placement
of the WiMax and WiFi mesh base stations on the UCLA campus.\textcolor{black}{\vspace{-3pt}}

\subsection{Collected Data\label{sec:Collected-data}}

\begin{figure*}
\subfloat[U/L, Packet Size: 512B]{\includegraphics[width=0.2\paperwidth]{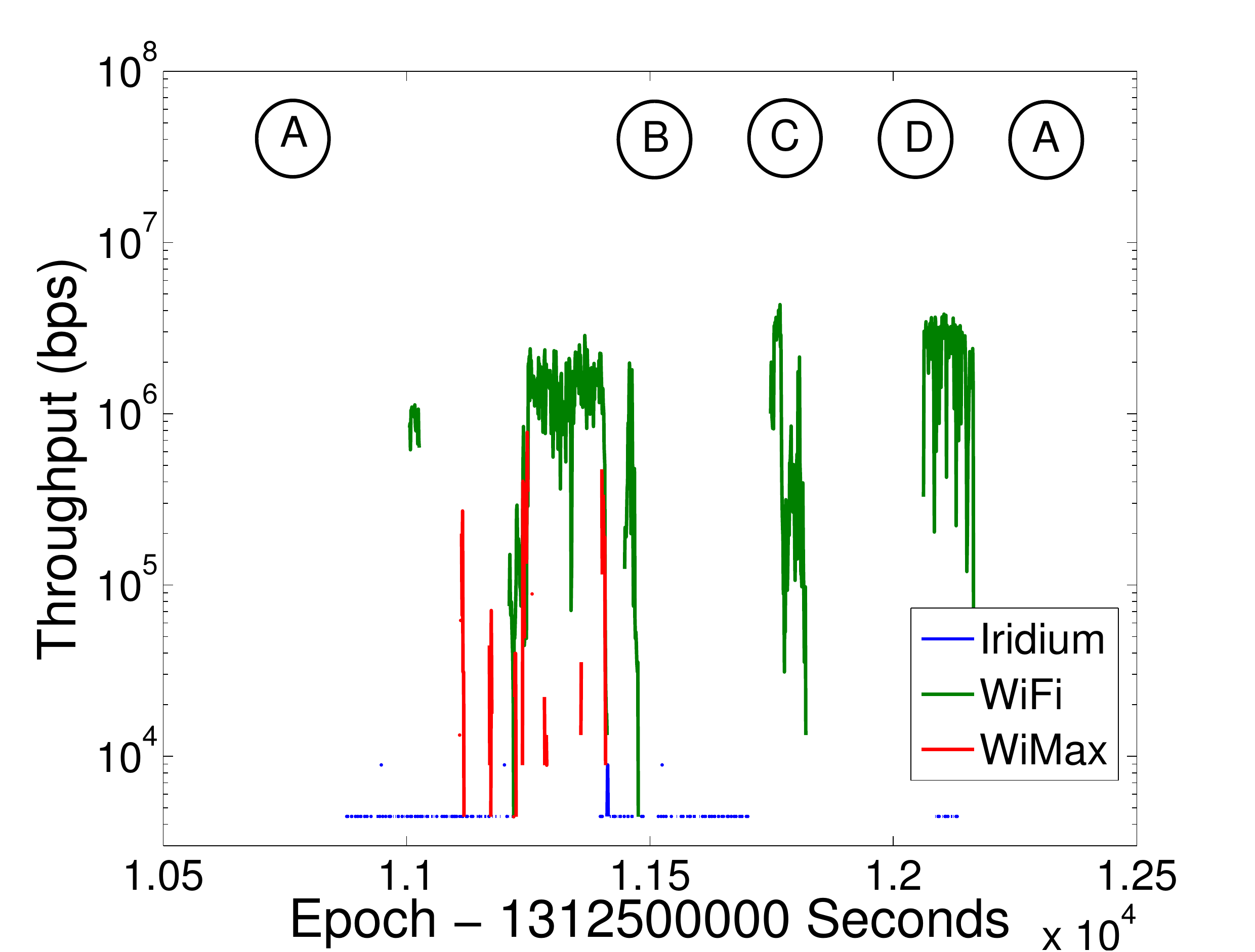}

}\subfloat[U/L, Packet Size: 1350B]{\includegraphics[width=0.2\paperwidth]{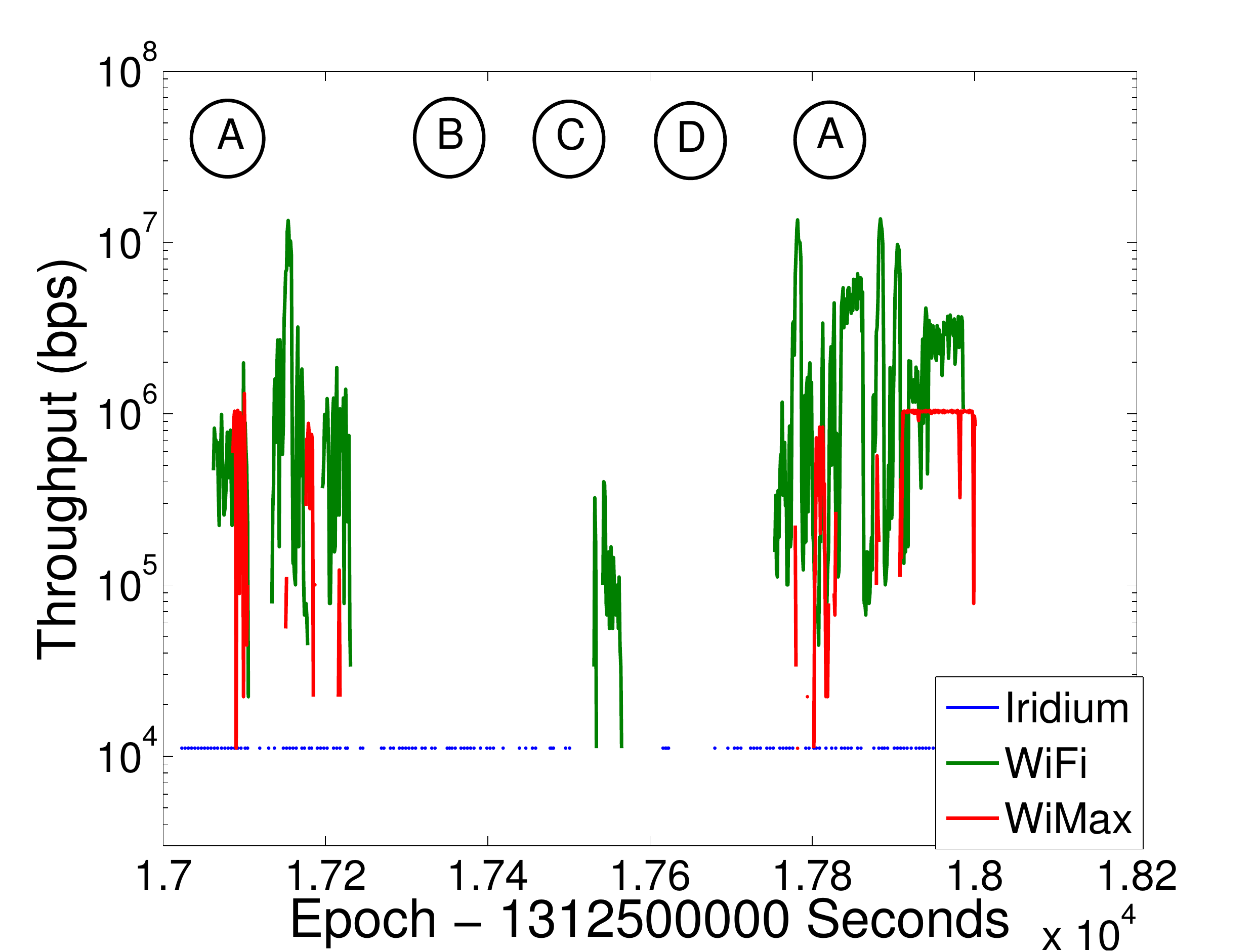}

}\subfloat[D/L, Packet Size: 64B]{\includegraphics[width=0.2\paperwidth]{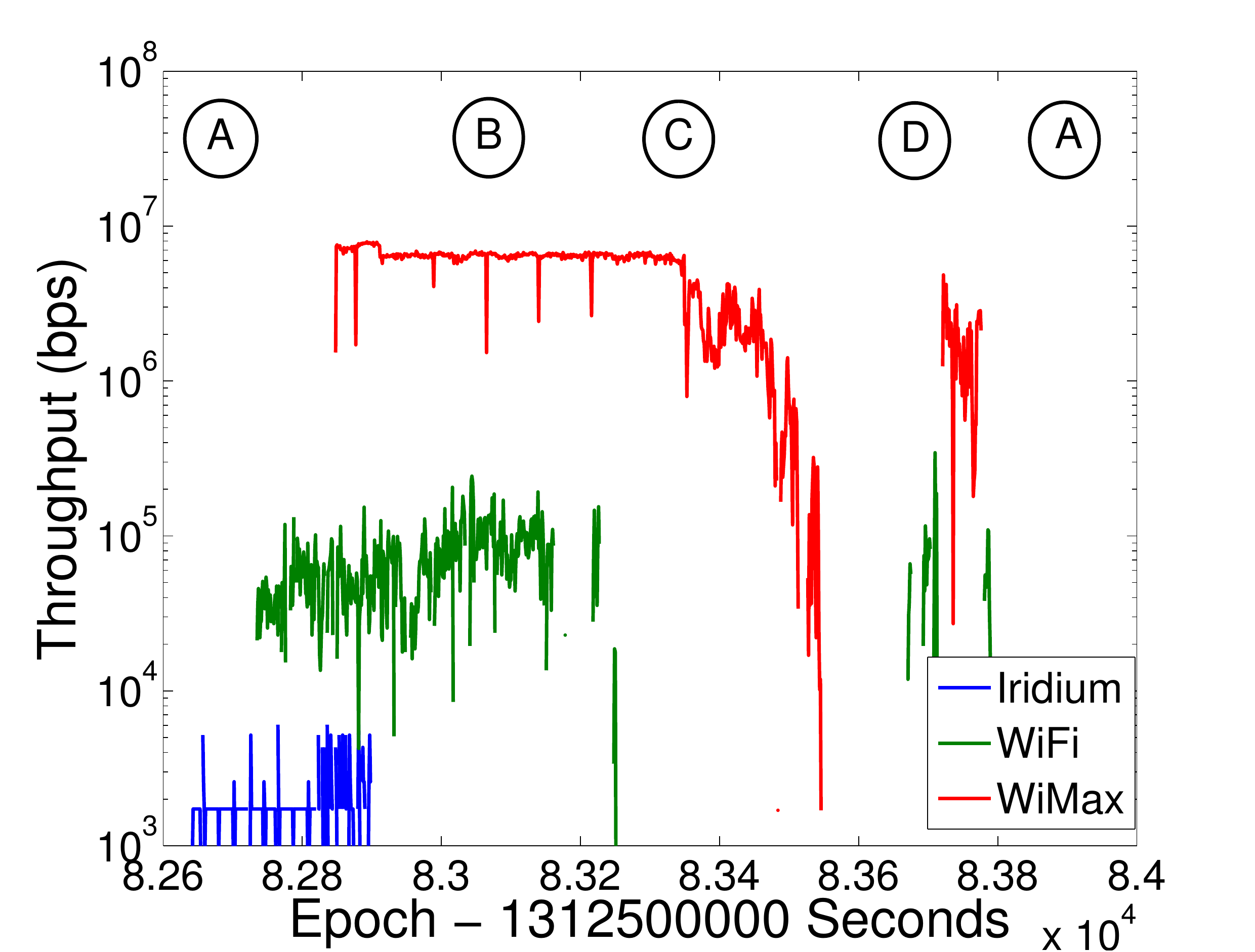}

}\subfloat[D/L, Packet Size: 1350B]{\includegraphics[width=0.2\paperwidth]{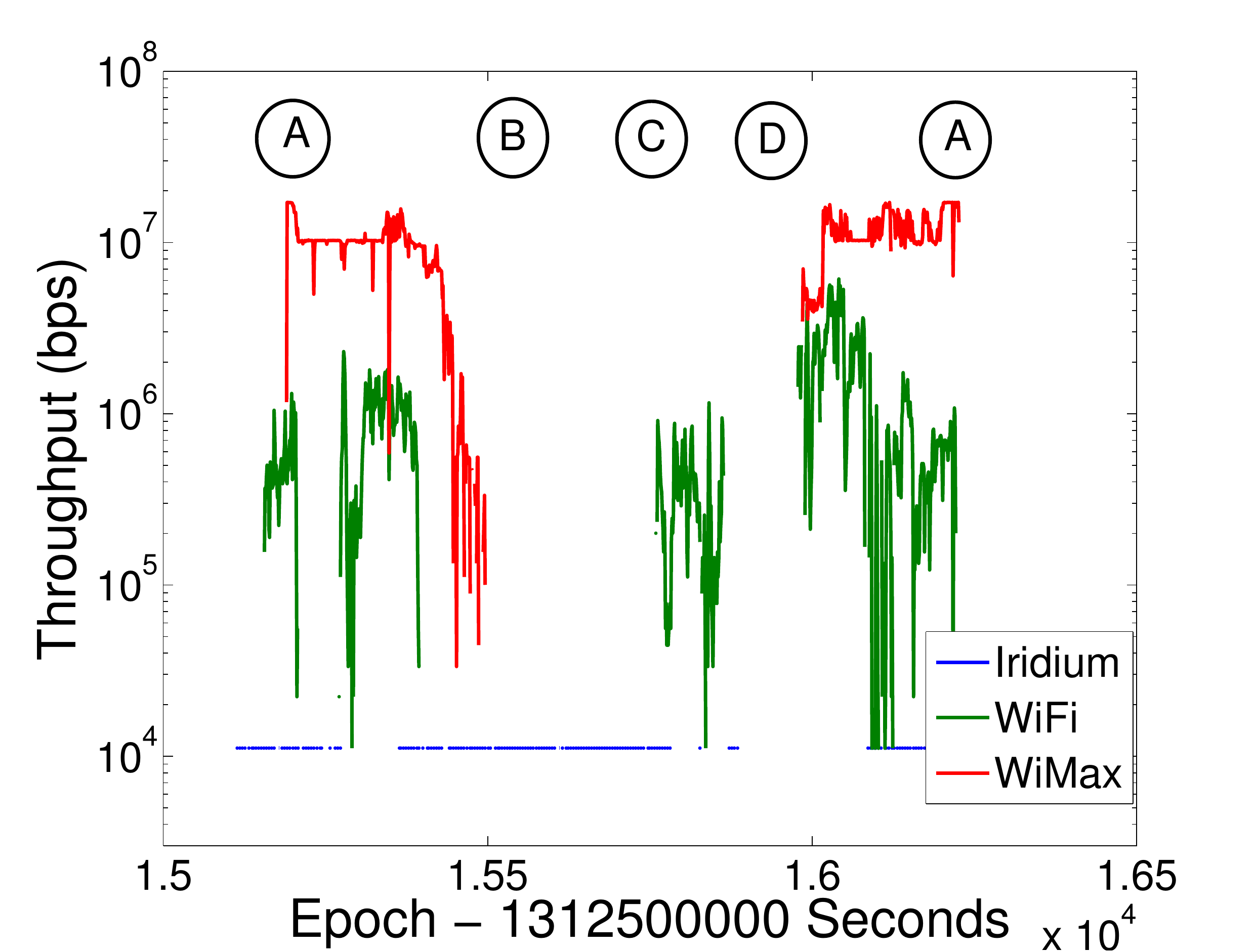}

}

\caption{Sample traces showing the UDP throughput for two U/L and two D/L experiments
with varying packet sizes. The labels A, B, C, and D provide the approximate
location of the vehicle when compared with Figure \ref{fig:Test-Setup}(b).
\label{fig:Traces}}
\textcolor{black}{\vspace{-20pt}}
\end{figure*}

Ten mobile experiments were conducted over a period of five days in
August 2011. For each of these experiments, traces were collected
and compared for each of the different networks. A sample of the collected
traces are shown in Figure \ref{fig:Traces}. These traces show the
UDP throughput for each network when all traffic is sent either to
(D/L) or from (U/L) the vehicle.

\begin{figure}
\begin{centering}
\subfloat[RTT CDF]{\begin{centering}
\includegraphics[width=0.5\columnwidth]{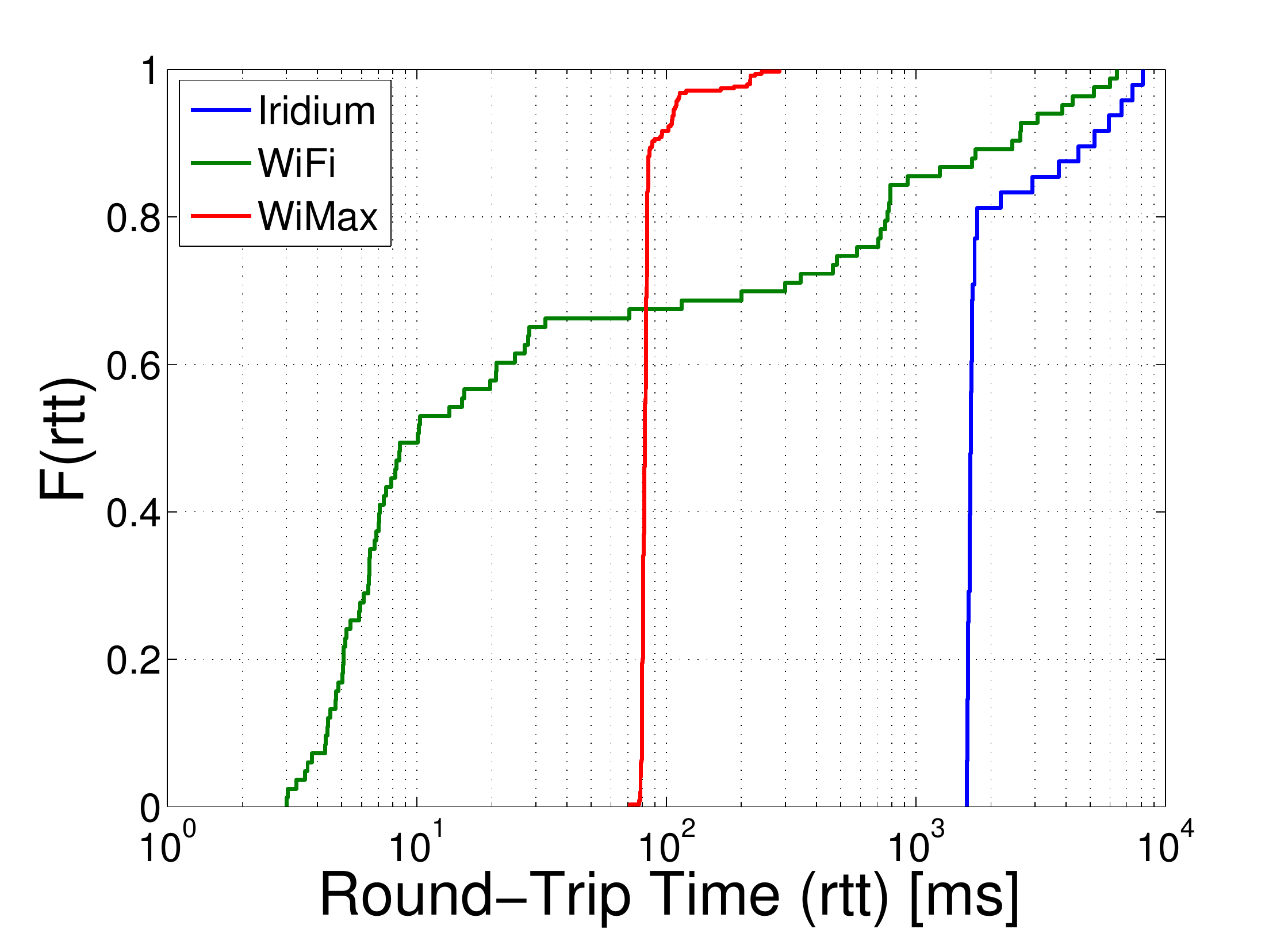}
\par\end{centering}

}\subfloat[Packet Loss CDF]{\begin{centering}
\includegraphics[width=0.5\columnwidth]{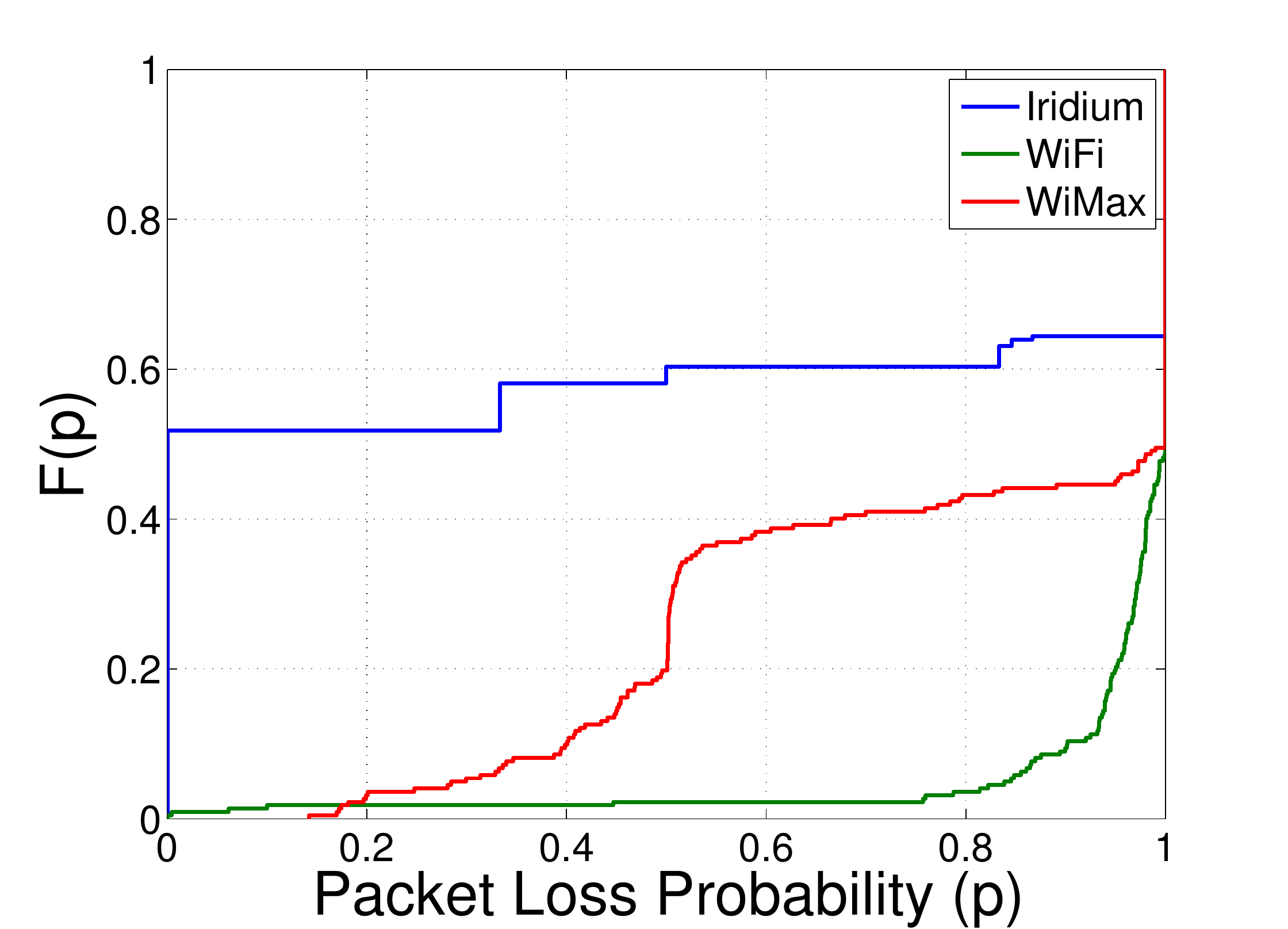}
\par\end{centering}

}
\par\end{centering}

\caption{CDFs of the RTT and packet loss probabilities during the D/L experiment
using 1,350 byte packets.\label{fig:Ping-CDF} }
\textcolor{black}{\vspace{-25pt}}

\end{figure}

The round-trip time (RTT) and packet loss probability for each network
was also collected. The CDFs for both the RTT and packet loss probability
during the D/L experiment where 1,350 byte packets were used is shown
in Figure \ref{fig:Ping-CDF}. The RTT was measured using ping messages
that were sent throughout the experiment on both the WiFi and WiMax
networks, while ping messages were only sent for approximately 60
seconds at the beginning of the experiment on the Iridium network
due to the bandwidth constraints of the network.\textcolor{black}{{}
The packet loss probabilities were determined by comparing the trace
files on both the NRL server and the vehicle computer.\vspace{-3pt}}

\subsection{Discussion and Comments on the Experimental Results\label{sub:Discussion}}

Data collected during each of the experiments provides information
about the expected environment that mobile users are likely to experience.
However, there are caveats concerning the methods in which the data
was collected that must be noted. First, the only user on each network
was the vehicle; although the WiFi mesh network performance was affected
by significant interference from adjacent WiFi networks and the Iridium
network was setup over an operational system. As a result, the WiMax
throughput presented in Figure \ref{fig:Traces} is much larger than
what would be expected when fully loaded, while the WiFi and Iridium
throughput is close to what we would expect in the real-world. The
RTT shown in Figure \ref{fig:Ping-CDF}(a) is also affected by this
situation. Since only one user has access to the WiMax network, the
RTT is fairly consistent throughout each of the experiments. The WiFi
RTT is largely affected by contention with adjacent WiFi networks
resulting in a wide range of possible RTTs, and the Iridium RTT is
consistent except for periods where we believe horizontal handoffs
between satellites occurred. Second, the data collection methods were
designed so that data could be used to replay each experiment off-line
enabling easy evaluation of future protocol designs. This prevented
us from collecting reliable statistics on the packet loss probabilities.
However, Figure \ref{fig:Ping-CDF}(b) shows the overall reliability
of each network and indicates that the satellite network provides
the most reliability and the WiFi network provides the least. Finally,
the use of a modem and the PSTN for the Iridium network (and consequently
the low throughput) is due to the Iridium system design. Iridium was
developed for world-wide voice communications. Modern satellite systems
do provide higher bandwidth, and therefore better performance for
packet based communication. Unfortunately, the use of these systems
was prohibitively expensive.

Regardless, the traces shown in Figure \ref{fig:Traces} indicate
that using a multi-path solution can potentially provide significant
performance gains over that of using only one of the networks exclusively.
Throughout each experiment, the vehicle was connected to at least
one network the majority of the time; and in many cases, it was connected
to two or more networks. Leveraging this connectivity can help ensure
that reliable, continuous data transport is an option in mobile environments.
The benefits of leveraging simultaneous networks for data transport
will be quantified in the following sections using the collected data.
Specifically, packet loss and RTT statistics will be used to provide
a comparison between the performance of MPTCP and MPTCP/NC in multi-path,
wireless scenarios.

\section{Analytical Models for Multi-Path TCP and Multi-Path TCP with Network
Coding\label{sec:Analytical-Models}}

\begin{figure}
\begin{centering}
\subfloat[MPTCP]{\begin{centering}
\includegraphics[width=0.4\columnwidth]{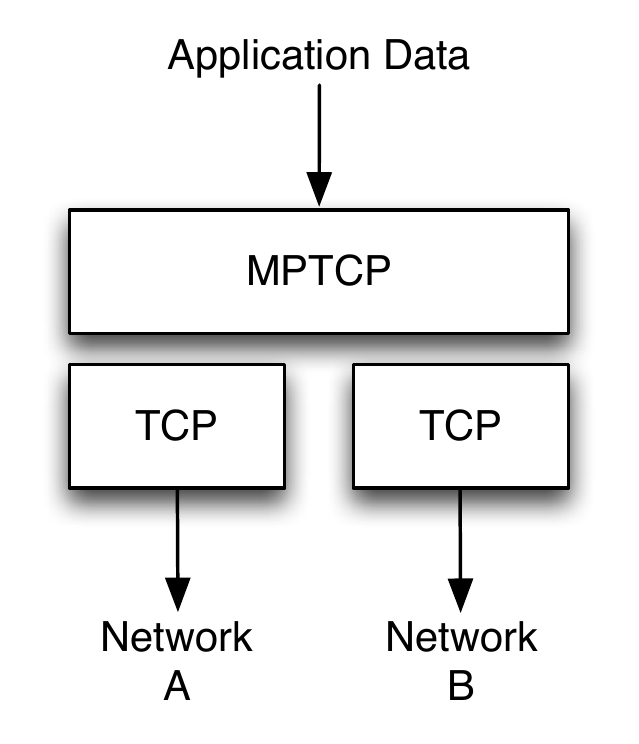}
\par\end{centering}

}\subfloat[MPTCP/NC]{\begin{centering}
\includegraphics[width=0.4\columnwidth]{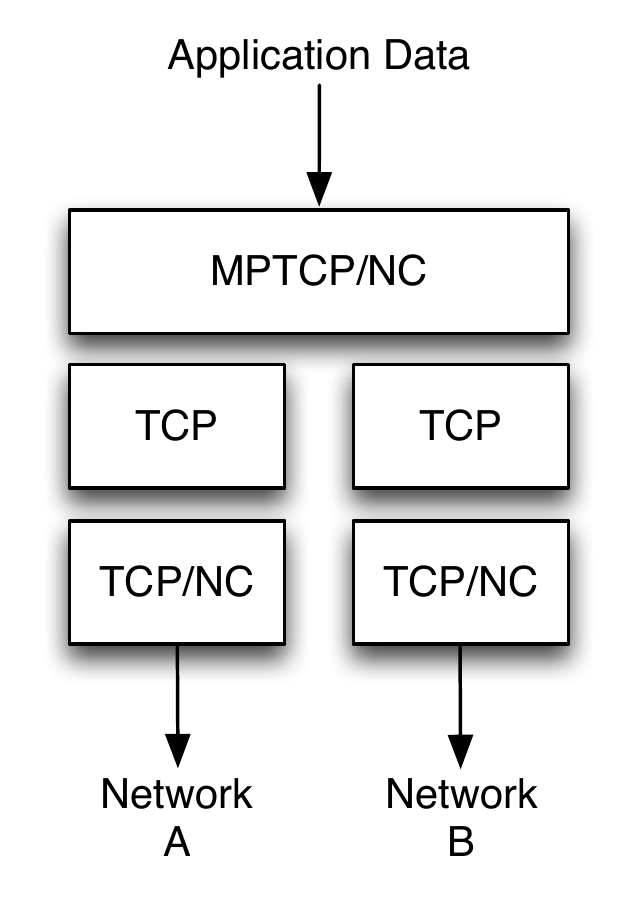}
\par\end{centering}

}\caption{Assumed network stack configuration for both MPTCP and MPTCP/NC.\label{fig:Assumed-network-stack}}
\textcolor{black}{\vspace{-25pt}}
\par\end{centering}

\end{figure}

Approaches similar to that of \cite{padhye2000modeling} and \cite{kim2011modeling}
are used to provide a mean-field approximation of the throughput for
both MPTCP and MPTCP/NC. The MPTCP analysis will assume the standard
implementation as shown in Figure \ref{fig:Assumed-network-stack}(a)
and defined by \cite{mptcpietf}. The MPTCP/NC analysis will assume
that the MPTCP/NC layer shown in Figure \ref{fig:Assumed-network-stack}(b)
provides a first layer of network coding before packets are injected
into a TCP sub-flow, and the TCP/NC layer provides a second layer
of network coding, similar to \cite{sundararajan2011network}, in
order to overcome random packet losses due to lossy networks.

The analysis for MPTCP will use the model presented by \cite{padhye2000modeling}
while assuming that perfect scheduling of packets across various TCP
sub-flows takes place. Once the analytical throughput for each individual
sub-flow is determined, the results can be summed to determine MPTCP's
overall throughput. In reality, perfect scheduling is not possible
due to packet losses, termination of a specific sub-flow, etc. This,
in-turn, results in the need to collect feedback regarding which packets
were lost, retransmit each lost packet on a second (or third) TCP
sub-flow, and verify receipt of that packet by the receiver. This
process significantly decreases the efficiency of MPTCP by both lowering
the throughput and increasing the transport time. With this in mind,
the analytical results presented later will over estimate the performance
of MPTCP.

In the case of MPTCP/NC, network coding can be used to aid in the
sub-flow scheduling problem by eliminating the need to track specific
packets sent over the network. With respect to the analysis, we will
assume that network coding is performed prior to a packet's injection
into a sub-flow. If the coding operations are carried out properly,
the receiver only needs to collect enough coded packets, or degrees
of freedom (DOF), in order to successfully transfer data over multiple
sub-flows without the need to track individual packets through the
multiple networks. Not only does this significantly decrease the complexity
of the protocol, but also provides greater freedom for determining
how to allocate packets among the collection of sub-flows. We will
also assume that a second layer of network coding occurs below TCP
and redundant packets are transmitted to overcome random packet losses.
In general, the number of transmitted packets for every DOF sent should
be $R\geq\nicefrac{1}{1-p}$ where $R$ is the redundancy and $p$
is the packet loss probability of the network path. \cite{sundararajan2011network}
provides a full description of the network coding operations and gains
that can be achieved using network coding in this manner.

Finally, we will assume that both protocols use a TCP Reno style of
congestion control on each sub-flow. This assumption keeps the results
presented here in line with those presented by \cite{padhye2000modeling}
and also simplifies the analysis for MPTCP/NC. Because we assume that
network coding is performed below TCP on each sub-flow, network coding
eliminates the need to consider the effects of triple-duplicates on
TCP's window size. A more detailed discussion will be provided in
subsequent sections.\textcolor{black}{\vspace{-5pt}}

\subsection{MPTCP Analytical Throughput}

The analytical throughput for MPTCP follows directly from \cite{padhye2000modeling}.
In \cite{padhye2000modeling}, the analytical throughput, $B(p)$,
of a single TCP connection was derived where $p$ is the independent
and identically distributed (i.i.d) loss probability of a single packet.
The equation for $B(p)$ can be found in equation (32) of \cite{padhye2000modeling}.
We extend this analysis to the multi-path case by taking the calculated
$B_{j}(p_{j})$ for each sub-flow, $j=\{1,\ldots,n\}$, and summing
them together to form the MPTCP throughput: 
\begin{equation}
B(p_{1},\ldots,p_{n})=\sum_{j=1}^{n}B_{j}(p_{j}).\label{eq:MPTCP-Throughput}
\end{equation}

As noted earlier, this does not let us take into account the inefficiencies
introduced by the MPTCP layer and will over-estimate the achievable
MPTCP throughput.\textcolor{black}{\vspace{-5pt}}

\subsection{Modeling MPTCP/NC's End-to-End Throughput}

\begin{figure}
\begin{centering}
\vspace{-10pt}\includegraphics[width=0.8\columnwidth]{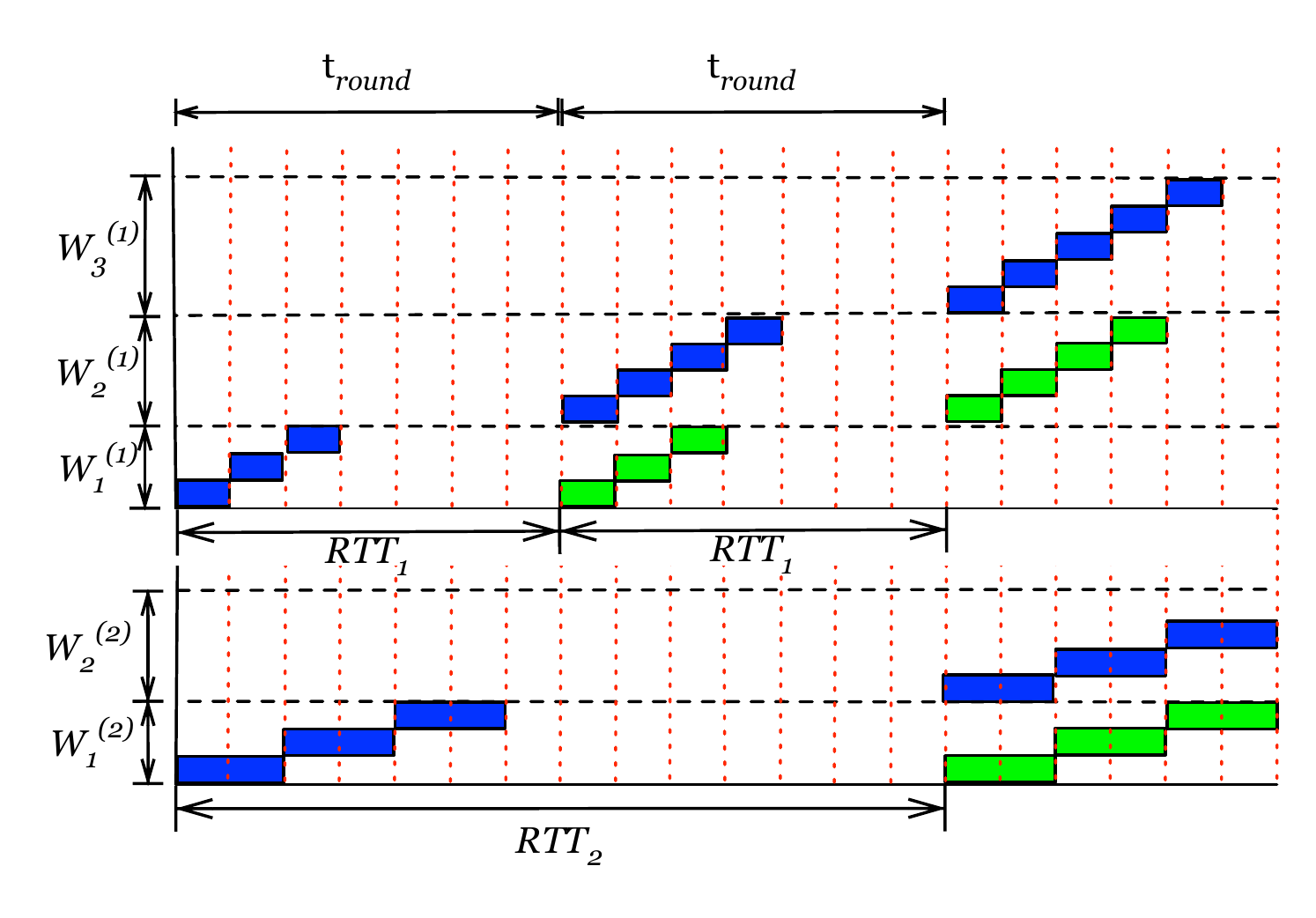}
\par\end{centering}

\caption{Round duration used for MPTCP/NC for two sub-flows. The blue blocks
indicate packets and the green blocks indicate acknowledgements.\label{fig:Round-duration}}
\textcolor{black}{\vspace{-20pt}}
\end{figure}

Two metrics will be used to develop the MPTCP/NC mean-field approximation:
the average throughput $\mathcal{T}$, and the expected MPTCP/NC congestion
window evolution $\mathbb{E}[W]$. We model MPTCP/NC's behavior in
terms of rounds. The natural choice for determining the duration of
a round is to use the RTT from the sender to the receiver (i.e., $t_{rnd}=RTT$).
While this works if there is a single TCP connection, each sub-flow
is expected to have different round trip times making it difficult
to determine which RTT to use. This is accounted for by setting the
duration of each round, $t_{rnd}$, equal to the greatest common divisor
(GCD) of the sub-flows' RTTs. Figure \ref{fig:Round-duration} provides
an illustration of this concept.

\subsubsection{MPTCP/NC Sub-Flow Analysis}

We now use the most basic implementation of TCP in our analysis and
initially assume that each round's duration is equal to the RTT of
sub-flow $j$. We assume that the congestion window size during round
$i$ is determined by the number of acknowledgements $a$ indicating
successfully transmitted packets obtained during round $i-1$: 
\begin{equation}
W_{i}^{(j)}=W_{i-1}^{(j)}+\frac{a_{i-1}^{(j)}}{W_{i-1}^{(j)}}.\label{eq:cwnd_window}
\end{equation}
This concept is also shown in Figure \ref{fig:Round-duration} where
the congestion window size of each sub-flow grows as a function of
the number of acknowledgements received. We now assume that $R_{j}$
linearly independent, redundant, network coded packets are sent for
each uncoded packet contained the TCP congestion window, there are
i.i.d. packet losses, and a packet loss rate of $p_{j}$. Taking the
expectation of the window size, $\mathbb{E}[W_{i}^{(j)}]$, we obtain:

\begin{eqnarray}
\mathbb{E}\left[W_{i}^{(j)}\right] & = & \mathbb{E}\left[W_{i-1}^{(j)}\right]+\textrm{min}\left(1,\left(1-p_{j}\right)R_{j}\right)\label{eq:Expected_W}\\
 & = & \mathbb{E}\left[W_{1}^{(j)}\right]+\left(i-1\right)\textrm{min}\left(1,\left(1-p_{j}\right)R_{j}\right),\label{eq:Iter_Expected_W}
\end{eqnarray}
where the minimization is required because the window size can only
increase by a maximum of one packet per round.

Since the throughput $\mathcal{T}_{i}^{(j)}$ per round is related
to the number of packets sent in that round, 
\begin{equation}
\mathcal{T}_{i}^{(j)}=\frac{\mathbb{E}\left[W_{i}^{(j)}\right]}{RTT_{j}}\textrm{min}\left(1,\left(1-p_{j}\right)R_{j}\right).\label{eq:Round_Throughput}
\end{equation}
The minimization in this equation is necessary to account for packets
that are received that do not deliver new degrees of freedom. Since
the TCP/NC layer codes all packets within the TCP congestion window,
delivered packets 1 through $W_{i}^{(j)}$ contain new degrees of
freedom. If more than $W_{i}^{(j)}$ packets are received in the round,
the MPTCP/NC layer disregards them since they contain no new information.

The above analysis assumed that the RTTs for each sub-flow was the
same. Because this is not necessarily the case, we must adjust equation
(\ref{eq:Iter_Expected_W}) to account for the shorter round durations
by defining $\alpha_{j}=\nicefrac{RTT_{j}}{t_{rnd}}$ and substituting
$\lceil\nicefrac{i}{\alpha_{j}}\rceil$ for $i$,

\begin{eqnarray}
\mathbb{E}\left[W_{\lceil\nicefrac{i}{\alpha_{j}}\rceil}^{(j)}\right] & = & \mathbb{E}\left[W_{1}^{(j)}\right]+\left(\lceil\nicefrac{i}{\alpha_{j}}\rceil-1\right)\textrm{min}\left(1,\left(1-p_{j}\right)R_{j}\right)\nonumber \\
 & = & \gamma_{i}^{(j)}.\label{eq:Exp_Window_Size_1}
\end{eqnarray}
The throughput for each TCP sub-flow $j$ then becomes $\mathcal{T}_{i}^{(j)}=\nicefrac{\gamma_{i}^{(j)}}{\left(\alpha_{j}\cdot t_{rnd}\right)}\min\left(1,\left(1-p_{j}\right)R_{j}\right)$,
which can be further reduced if we consider a large enough redundancy
factor $R_{j}$. For $R_{j}>\nicefrac{1}{\left(1-p_{j}\right)}$,
the instantaneous throughput becomes, 
\begin{equation}
\mathcal{T}_{i}^{(j)}=\frac{1}{\alpha_{j}\cdot t_{rnd}}\left(\mathbb{E}\left[W_{1}^{(j)}\right]+\lceil\nicefrac{i}{\alpha_{j}}\rceil-1\right).
\end{equation}

Finally, we account for the fact that the number of packets sent in
each RTT is upper-bounded by TCP's maximum congestion window size,
$W_{\text{max}}^{(j)}$. This results in:
\begin{equation}
\mathcal{T}_{i}^{(j)}=\frac{1}{\alpha_{j}\cdot t_{rnd}}\left(\textrm{min}\left(W_{\textrm{max}}^{(j)},\mathbb{E}\left[W_{1}^{(j)}\right]+\lceil\nicefrac{i}{\alpha_{j}}\rceil-1\right)\right).
\end{equation}

The model we used in our analysis of the MPTCP/NC sub-flow performance
makes several assumptions that, in practice, should be considered.
First, we assume that packet losses are i.i.d. with loss probability
$p_{i}$. Therefore, the analysis does not account for correlated
packet losses due to congestion and other factors. Second, we assumed
that $R_{j}$ is sufficiently large enough to ignore the possibility
of time-outs. While the probability of a time-out decreases with increasing
$R_{j}$, time-outs still occur in practice and the impact of each
time-out on the throughput is significant (i.e., the congestion window
size is reset to $\mathbb{E}[W_{1}^{(j)}]$). Specifically, a time-out
occurs when the sum of received acknowledgements over two rounds,
$i$ and $i+1$, is less than the window size during round $i$ with
probability $\mathcal{P}r\left(a_{i}+a_{i+1}<W_{i}\right)$. Generalizing
equation (\ref{eq:cwnd_window}) to account for time-outs, with respect
to $W_{\text{max}}$ and $R_{j}$, may allow for a bound on the decrease
in throughput to be determined resulting in a more accurate approximation.
Third, we assume that $RTT_{j}$ remains constant. In practice, this
is not true, and implementations of TCP generally use an averaged
round-trip time often referred to as the ``smoothed'' round-trip
time $SRTT$.

\subsubsection{MPTCP/NC's Window Evolution and End-to-End Throughput}

Using the above results, the average end-to-end MPTCP/NC throughput
over $k$ rounds is determined using a round duration of $t_{rnd}$
and defining $\alpha_{j}=\nicefrac{RTT_{j}}{t_{rnd}}$,

\begin{equation}
\mathcal{T}\left(k\right)=\frac{1}{k}\sum_{i=1}^{k}\mathcal{T}_{i}=\frac{1}{k}\sum_{i=1}^{k}\sum_{j=1}^{n}\mathcal{T}_{i}^{(j)}.
\end{equation}
Assuming that $\nicefrac{k}{\alpha_{j}}\in\mathbb{Z}$, $\gamma_{i}^{(j)}\leq W_{max}^{(j)}$,
and relaxing $\lceil\nicefrac{i}{\alpha_{j}}\rceil$ so that it is
$\nicefrac{i}{\alpha_{j}}$ for all $j$,
\begin{eqnarray}
\mathcal{T}(k) & = & \frac{1}{k}\sum_{j=1}^{n}\left(\frac{1}{\alpha_{j}\cdot t_{rnd}}\sum_{i=1}^{k}\gamma_{i}^{(j)}\right)\\
 & = & \frac{1}{t_{rnd}}\sum_{j=1}^{n}\left(\frac{1}{\alpha_{j}}\mathbb{E}\left[W_{1}^{(j)}\right]+\frac{k+1}{2\alpha_{j}^{2}}-\frac{1}{\alpha_{j}}\right),\label{eq:unbounded-tp}
\end{eqnarray}
If $\nicefrac{k}{\alpha_{j}}\notin\mathbb{Z},\forall j$, the above
equation will contain additional terms that contain packets sent in
the rounds from $\lfloor\nicefrac{k}{\alpha_{j}}\rfloor$ to $\nicefrac{k}{\alpha_{j}}$.
Furthermore, the relaxation of $\lceil\nicefrac{i}{\alpha_{j}}\rceil$
to $\nicefrac{i}{\alpha_{j}}$ decreases the throughput since we are
no longer accounting for $\lceil\nicefrac{i}{\alpha_{j}}\rceil-\nicefrac{i}{\alpha_{j}}$
packets sent per round. As $k$ grows, these approximations have less
of an effect on the throughput.

Finally, we take into account the maximum window size of each sub-flow
$W_{\textrm{max}}^{(j)}$; but first, we define: 
\begin{equation}
r^{(j)}=\alpha_{j}\left(W_{\textrm{max}}^{(j)}-\mathbb{E}\left[W_{1}^{(j)}\right]\right).
\end{equation}
 Using equation (\ref{eq:unbounded-tp}) and assuming that $R_{j}>\nicefrac{1}{(1-p_{j})}$,
the average end-to-end throughput ${\cal T}_{e2e}$, in packets per
second is:
\begin{equation}
\mathcal{T}_{e2e}(k)=\sum_{j=1}^{n}{\cal T}_{e2e}^{(j)}(k),\label{eq:sum_mptcpnc_tp}
\end{equation}
where
\begin{equation}
{\cal T}_{e2e}^{(j)}(k)=\begin{cases}
\frac{1}{\alpha_{j}\cdot t_{rnd}}\left(\mathbb{E}\left[W_{1}^{(j)}\right]+\frac{k+1}{2\alpha_{j}}-1\right) & \textrm{for }k\leq r^{(j)},\\
\frac{\rho^{(j)}}{\alpha_{j}\cdot k\cdot t_{rnd}} & \textrm{for }k>r^{(j)},
\end{cases}\label{eq:mptcp-tp}
\end{equation}
and
\begin{equation}
\rho^{(j)}=r^{(j)}\mathbb{E}\left[W_{1}^{(j)}\right]+\frac{r^{(j)}\left(r^{(j)}+1-2\alpha_{j}\right)}{2\alpha_{j}}+W_{\text{max}}^{(j)}\left(k-r^{(j)}\right).
\end{equation}

It should be noted that as $k\rightarrow\infty$ for $R_{j}>\nicefrac{1}{(1-p_{j})}$,
the average end-to-end throughput $T_{\text{e2e}}(k)\rightarrow\sum_{j=1}^{n}\nicefrac{W_{\text{max}}^{(j)}}{\left(\alpha_{j}\cdot t_{tnd}\right)}$.

\section{Comparison of Multi-Path TCP and Multi-Path TCP with Network Coding
using Empirical Data\label{sec:Performance}}

\begin{figure}
\subfloat[D/L, Packet Size: 1350B]{\begin{centering}
\includegraphics[width=0.5\columnwidth]{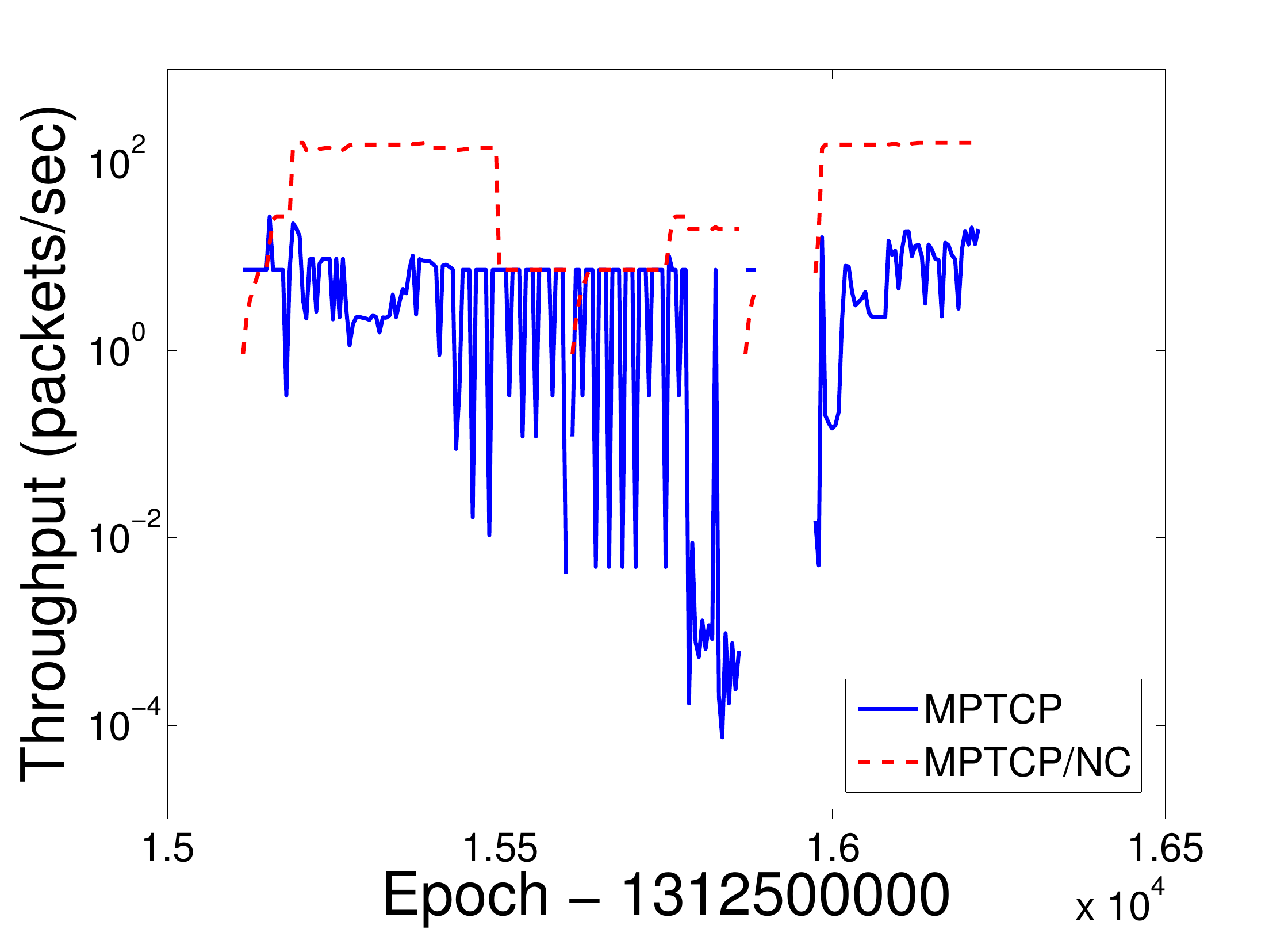}
\par\end{centering}

}\subfloat[U/L, Packet Size: 1350B]{\begin{centering}
\includegraphics[width=0.5\columnwidth]{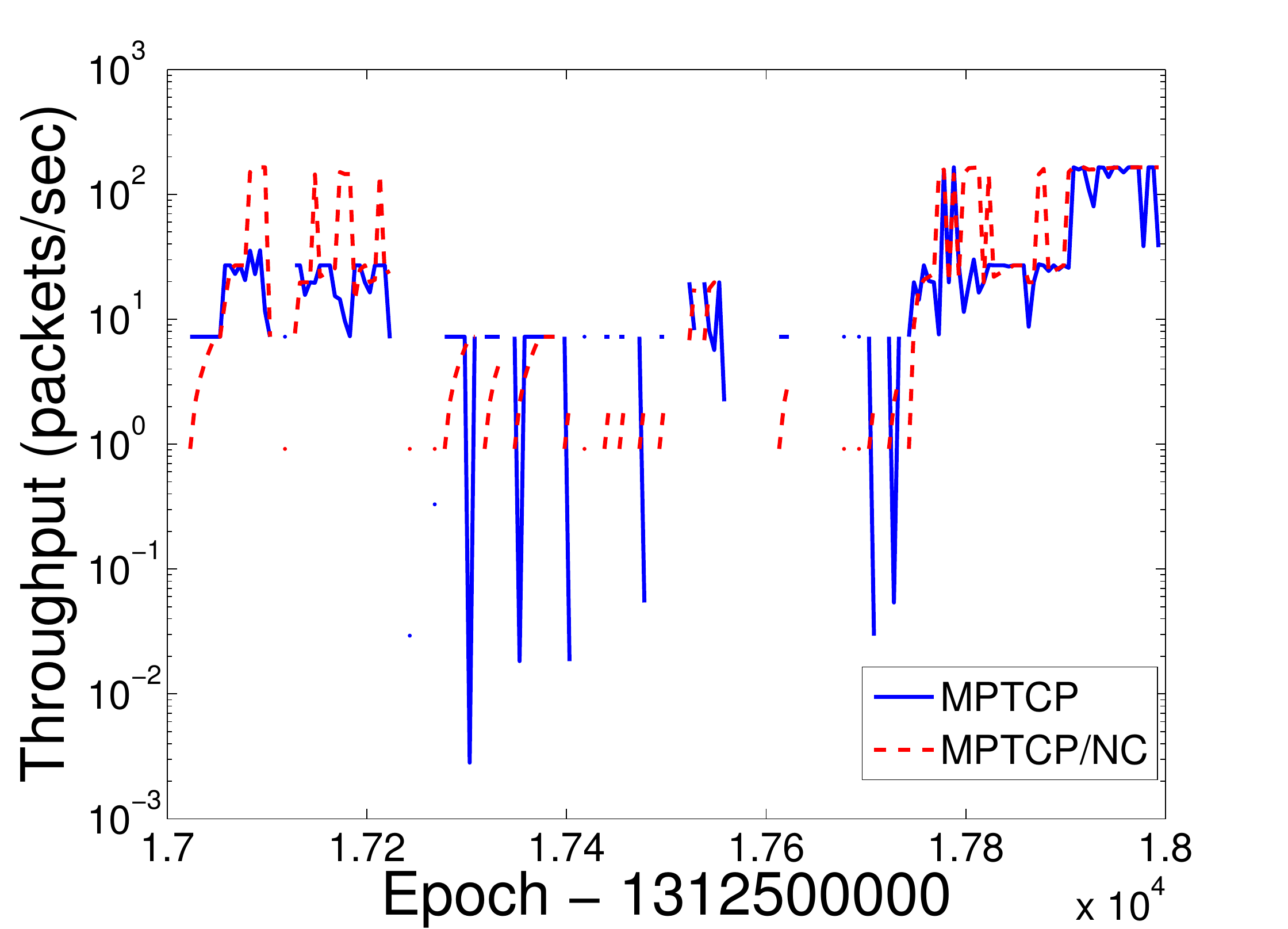}
\par\end{centering}

}

\caption{Comparison of the theoretical MPTCP and MPTCP/NC throughput using
the data presented in Section \ref{sec:Empirical-measurements}.\label{fig:Comparison-of-theoretical}}
\textcolor{black}{\vspace{-25pt}}
\end{figure}

We now compare the theoretical throughput of MPTCP, equation (\ref{eq:MPTCP-Throughput}),
with that of the theoretical throughput of MPTCP/NC, equation (\ref{eq:sum_mptcpnc_tp}).
Figure \ref{fig:Comparison-of-theoretical} shows the performance
of both protocols using the data presented in Section \ref{sec:Empirical-measurements}
as a baseline. The maximum window size for each TCP sub-flow was set
to $W_{\text{max}}=12$; and a mean RTT, based off of empirical data,
was used for each network where $RTT_{\text{Iridium}}=1.653\text{s}$,
$RTT_{\text{WiFi}}=0.607\text{s}$, and $RTT_{\text{WiMax}}=0.087\text{s}$.
Empirical packet loss data, averaged over 5s, on each separate path
from two of the experiments was used as a baseline for determining
both throughputs. It was assumed that the network capacity for each
network was large enough to send $W_{\text{max}}$ packets in the
case of MPTCP and $R_{j}W_{\text{max}}$ packets in the case of MPTCP/NC
where $R_{j}$ is assumed to be large enough so that time-outs are
very unlikely (i.e., $R_{j}$ is much larger than the 5s average of
the packet loss probability). In addition, Figure \ref{fig:Comparison-of-theoretical}
uses the mean-field approximations developed in the last section and
does not show a simulated behavior of each protocol.

The figures show that MPTCP/NC provides a better throughput throughout
the ``simulated'' experiment than MPTCP. While MPTCP is severely
hindered by high packet losses as a result of poor channel conditions,
MPTCP/NC is able to mask the majority of packet losses and maintain
a high throughput. Scheduling of packets on each sub-flow is also
easier with MPTCP/NC than with MPTCP due to the network coding operations
performed immediately below the application layer. The throughput
shown for MPTCP assumes that there is perfect scheduling among the
sub-flows with no need to retransmit a packet on more than one sub-flow.
This provides a best-case scenario for the achievable throughput.
This assumption is not made in MPTCP/NC because each packet transmitted
on a sub-flow is viewed as a degree of freedom. If a packet is lost,
any packet sent on a different sub-flow can be used in the lost packet's
place.

\section{Conclusion\label{sec:Conclusion}}

We have presented empirical measurements for the simultaneous use
of three heterogeneous networks showing that the combined use of all
three is needed in order to provide an improved level of performance
in mobile environments. We then suggested the use of a multi-path
protocol based on MPTCP that uses network coding to overcome the challenges
of packet scheduling and lossy wireless networks. A mean-field approximation
of the throughput for both MPTCP and MPTCP/NC was developed and used,
along with the empirical data, to provide a comparison of the two
protocols. This comparison showed that the use of network coding in
multi-path, lossy scenarios can significantly increase the quality
of service for mobile users.

\section*{Acknowledgements\label{sec:Acknowledgements}}

{\small A large team was required to collect and analyze the experimental
data and the authors would like to acknowledge their contribution.
Giovanni Pau and Mario Gerla from UCLA, as well as Au Teerapittayanon
and} Danail {\small Traskov from MIT, were critical to the success
of the data collection efforts. Jamie Simons from MIT was also critical
to the data processing efforts. This paper would not be possible if
it was not for their help.}{\small \par}

\bibliographystyle{ieeetr}
\bibliography{Network_Coding_over_Multiple_Heterogeneous_Networks_in_Mobile_Environments_v2}

\end{document}